\documentclass[aps,prb,twocolumn,superscriptaddress,showpacs]{revtex4-2}
\usepackage{amsmath,amsfonts, amssymb, dsfont}
\usepackage{yfonts}
\usepackage{bm}
\usepackage{graphicx}
\usepackage{verbatim}
\usepackage{hyperref}
\usepackage{xcolor}
\usepackage{enumerate}
\usepackage{tikz}
\usepackage{bbold}
\usepackage{booktabs}
\usepackage{multirow}

\usepackage[utf8]{inputenc}
\usepackage[T1]{fontenc}
\usepackage{xcolor}

\usepackage{diagbox}
\usepackage{tabularx}

\IfFileExists{newtxtext.sty}
{\usepackage{newtxtext,newtxmath}}
{\IfFileExists{stix.sty}
{\usepackage{stix}}
{\IfFileExists{mathptmx.sty}
{\usepackage{mathptmx}}{} } }

\usepackage{textcomp}
\usepackage{bm}

\usepackage{booktabs,siunitx}

\pdfoutput=1
\usepackage{color}
\definecolor{LinkColor}{rgb}{0.256,0.439,0.588}
\usepackage{hyperref}
\hypersetup{
colorlinks=true,
citecolor=LinkColor,
linkcolor=LinkColor,
urlcolor=LinkColor
}

\usepackage{multirow}

\begin{document}
%\linenumbers 
\title{Incremental SWAP Operator for Entanglement Entropy: Application for Exponential Observables in Quantum Monte Carlo Simulation}

\author{Xuan Zhou}
\affiliation{State Key Laboratory of Surface Physics, Fudan University, Shanghai 200438, China}
\affiliation{Center for Field Theory and Particle Physics, Department of Physics, Fudan University, Shanghai 200433, China}

\author{Zi Yang Meng}
%\email{zymeng@hku.hk}
\affiliation{Department of Physics and HKU-UCAS Joint Institute of Theoretical and Computational Physics, The University of Hong Kong, Pokfulam Road, Hong Kong SAR, China}

\author{Yang Qi}
\email{qiyang@fudan.edu.cn}
\affiliation{State Key Laboratory of Surface Physics, Fudan University, Shanghai 200438, China}
\affiliation{Center for Field Theory and Particle Physics, Department of Physics, Fudan University, Shanghai 200433, China}
\affiliation{Collaborative Innovation Center of Advanced Microstructures, Nanjing 210093, China}

\author{Yuan Da Liao}
\email{ydliao@hku.hk}
\affiliation{State Key Laboratory of Surface Physics, Fudan University, Shanghai 200438, China}
\affiliation{Center for Field Theory and Particle Physics, Department of Physics, Fudan University, Shanghai 200433, China}
\affiliation{Department of Physics and HKU-UCAS Joint Institute of Theoretical and Computational Physics, The University of Hong Kong, Pokfulam Road, Hong Kong SAR, China}
\date{\today}

\begin{abstract}
We propose a new method to efficiently compute the entanglement entropy (EE) of quantum many-body systems. Our approach, called the {\it incremental SWAP operator method}, combines the simplicity of the SWAP operator used in projector quantum Monte Carlo simulations with recent advances in precisely computing exponential observables using incremental algorithms. We apply this technique to obtain accurate EE data at reduced computational cost for 1d and 2d antiferromagnetic Heisenberg models with different bipartition schemes.
Using the computed EE data, we extract the area law coefficient, universal logarithmic corrections from Goldstone modes, and the geometric constant, finding quantitative agreement with analytical predictions. Moreover, for the first time in an unbiased numerical simulation of 2d antiferromagnetic Heisenberg model, we successfully obtain reliable universal logarithmic corrections from sharp corners that match expected theoretical values.
The consistency between our numerical results and theoretical calculations demonstrates the power of our approach for accessing challenging universal entanglement properties. 
The extensions of our method to other quantum spin/boson models and the interacting fermion models, are outlined.
\end{abstract}

\maketitle

\section{Introduction} 
Entanglement entropy (EE) encapsulates universal properties of quantum ground states due to its non-local nature and serves as a pivotal tool for investigating quantum many-body systems~\cite{calabreseEntanglement2004,eisertColloquium2010,laflorencieQuantum2016} . Recently, researchers have initiated inquiries into universal properties at quantum critical points (QCPs) using EE from large-scale quantum Monte Carlo (QMC) calculations in bosonic and fermionic systems~\cite{demidioUniversal2022,daliaoteaching2023,liuDisorder2023,zhaoScaling2022,zhaoMeasuring2022,songDeconfined2023,songExtracting2023,songResummation-based2023}, highlighting the growing awareness of the significance of EE in extracting fundamental properties of strongly correlated systems.
Theoretically, it is known that in conformal field theories (CFTs) of (1+1)d, the leading contribution to EE shows a logarithmic scaling with the size of the subregion, with coefficient proportional to the central charge ~\cite{holzheyGeometric1994}. In CFTs at higher dimensions, however, it is the subleading terms carry the universal information, while the EE data are dominated by the non-universal area law term~\cite{casiniUniversal2007,buenoUniversality2015}. For example, at (2+1)d QCPs, when the boundary of a subregion $A$ has corners with angles $\alpha_i$, one expects the 2nd R\'enyi EE to follow a relation~\cite{songExtracting2023}
\begin{equation}
    S_A^{(2)}(L) = a L + s_{c}\ln(L) + O(1/L),
\label{eq:eq1}
\end{equation}
where $a$ is a non-universal area law coefficient, $L$ is the boundary length of $A$, and the logarithmic correction coefficient $s_{c}$ can be expressed as a sum of contributions from each corner angle $\alpha_i$: $s_{c}= \sum_i s(\alpha_i)$. When $\alpha$ is close to $\pi$ (no sharp corners),  it is expected to $s(\alpha) \sim \sigma (\pi-\alpha)^2$ and $\sigma$ is proportional to the stress tensor central charge of the CFT~\cite{buenoUniversality2015,faulknerShape2015}.

The EE can also have a universal subleading contribution in an ordered phase that spontaneously breaks a continuous symmetry, i.e., with Goldstone modes~\cite{kallinAnomalies2011,metlitskiEntanglement2011}, such as in the 2d antiferromagnets 
\begin{equation}
S_A^{(2)} (L) =a L + s_G\ln(\frac{\rho_s}{v} L)+s_{c}\ln(L) + \gamma_{\text{ord}},
\label{eq:eq2}
\end{equation}
where $\rho_s$ denotes the spin stiffness at thermodynamic limit, $v$ represents the velocity of Goldstone modes, and $\gamma_{\text{ord}}$ is a universal geometric constant depending on the partition scheme~\footnote{Here, we choose to explicitly include $v$ and $\rho_s$ because their finite-size corrections are needed in the fitting in Eq.~\eqref{eq:s2-stripe-deng}.}.
Especially, the logarithmic correction contains two distinct contributions. $s_G= \frac{n_G}{2}$ is independent of the shape of $A$ and counts the number of Goldstone modes, which has been numerically computed with success in  Refs.~\cite{kulchytskyyDetecting2015,zhaoMeasuring2022,dengImproved2023,songExtracting2023}. The other term, $s_{c}$ is the corner correction similar with that in CFT, and it is much smaller in amplitude than $s_G$ (with the opposite sign), its accurate extraction is more challenging, with a few less accurate numerical attempts~\cite{KallinEntanglement2013,kallinCorner2014,HelmesEEbilayer2014,helmesUniversal2016}.

To obtain the EE of quantum many-body systems, unbiased numerical techniques play a crucial role. Exact diagonalization and density matrix renormalization group methods can directly measure ground state wavefunctions and accurately determine EE, but these methods are limited by small system sizes, precisely because the exponential growth of the Hilbert space and the fact that the EE is one of the exponential observable~\cite{daliaoControllable2023,zhangIntegral2023}. One needs a method that could control the exponential growth of coefficient of variation of EE, in polynomial computational complexity, and this is exactly in the spirit of the Monte Carlo simulation, by imposing the important sampling according to the probability distribution of the target observable. 
The real question here, is then, how to implement such important sampling in the QMC computation of EE.

In fact, early QMC attempts have already been applied to calculate the 2nd R\'enyi EE in 2d quantum models, mainly by the SWAP operator -- an instance of the reduced density matrix for a QMC configuration -- based algorithm and its variations~\cite{hastingsMeasuring2010,melkoFinitesize2010a,kallinAnomalies2011,inglisEntanglement2013,inglisWangLandau2013,humeniukQuantum2012,luitzImproving2014}. 
However, directly applying these QMC methods to calculate EE has been shown to have serious issue with convergence~\cite{hastingsMeasuring2010} (we refer to as the non-convergence problem of SWAP operator hereafter).
The underlying cause of this problem has not been clearly recognized or identified.
Recently, with the fast developments of the incremental algorithm~\cite{albaOutofequilibrium2017,demidioEntanglement2020}, the non-convergence problem of the EE computation has been greatly reduced (although still not in the SWAP operator calculation), the more reliable data to extract the universal properties of EE from large-scale QMC calculations in 2d bosonic and fermionic system have been obtained~\cite{demidioUniversal2022,panStable2023,daliaoteaching2023,daliaoControllable2023,liuDisorder2023,zhaoScaling2022,zhaoMeasuring2022,songQuantum2023,songDeconfined2023,songResummation2023,zhangIntegral2023,songExtracting2023}, and the true reason behind the non-convergence problem begins to be understood.

In this paper, by combining the simplicity of the SWAP operator in the projector quantum Monte Carlo (PQMC) simulation~\cite{hastingsMeasuring2010,kallinAnomalies2011} and the recent understanding of computation of the exponential observables in the incremental algorithm~\cite{daliaoControllable2023,zhangIntegral2023}, we quantitatively demonstrate that the origin of the non-convergence problem of previous EE calculation lies in the fact that EE is an exponential observable, with exponential explosion of coefficient of variation as $L$ increases. We further provide a solution to this problem, dubbed {\it{incremental SWAP operator}} method with improved {\it propagating update} scheme for PQMC, to obtain the 2nd R\'enyi EE with less computational cost and more accurate data.
We apply the method upon 1d and 2d antiferromagnetic Heisenberg models and obtain results with better quality than the simple SWAP results, and in the 2d case, we successfully obtained the area law coefficient, the universal logarithmic corrections, both $s_G$ and $s_c$, and the the geometric constant $\gamma_{\text{ord}}$ in Eq.~\eqref{eq:eq2}, all of which, are well consistent with the theoretical predictions. We note that previous QMC results have never obtained all the coefficients accurately as we have done here. The extensions of our method to other quantum spin/boson models and the interacting fermion models, are straightforward.

\begin{figure}[htp!]
	\centering
	\includegraphics[width=1.0\columnwidth]{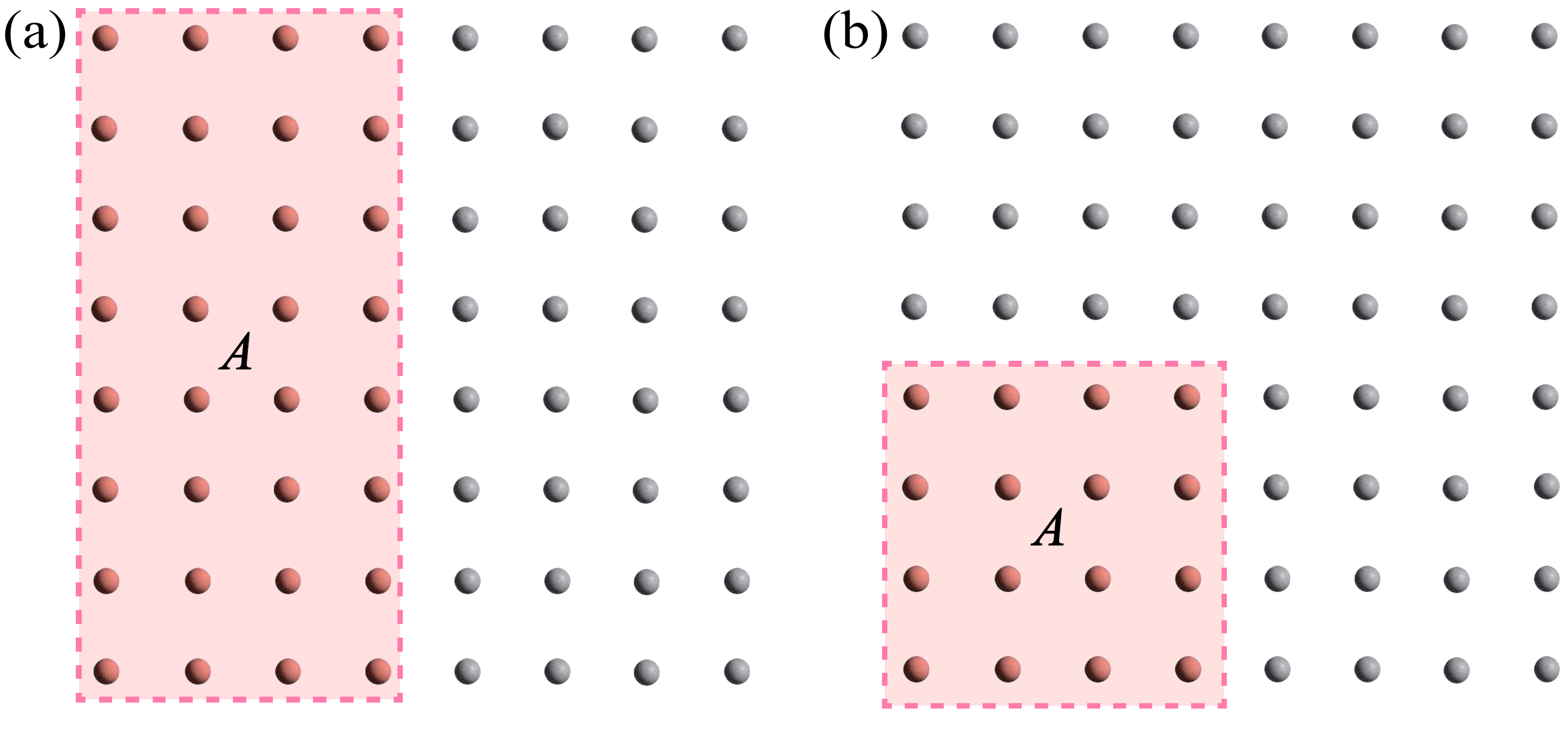}
	\caption{\textbf{Bipartition geometries for EE computation.} (a) $L/2 \times L$ stripe region $A$  with smooth boundaries. (b) $L/2\times L/2$ square entanglement region $A$  with four $\pi/2$ corners. The lattice is periodic in both dimensions.}
	\label{fg:fig1}
\end{figure}

\section{Non-convergence Problem for EE computation}
The numerical definition for the 2nd R\'enyi EE by SWAP operator in PQMC with valence-bond (VB) basis is given in Ref.~\cite{hastingsMeasuring2010} as $S_A^{(2)} = -\ln \langle \text{SWAP}_A\rangle$, with the observable
\begin{equation}\label{eq:swap}
\langle \text{SWAP}_A \rangle = \frac{ \sum\limits_{lr} w_{l} w_{r}\left\langle V_{l}^0\left|(-H)^m \text{SWAP}_A   (-H)^m\right| V_r^0 \right\rangle}{  \sum\limits_{lr} w_{l} w_{r} \left\langle  V_{l}^0 \left|(-H)^{2 m}\right|V_{r}^0 \right\rangle},
\end{equation}
where $\left|V_l^0\right\rangle$,$\left|V_r^0\right\rangle$ are arbitrary VB states with Kronecker product of two independent replica of Hamiltonian, $\sum_l w_l\left|V_l^0\right\rangle$ and $\sum_r w_r\left|V_r^0\right\rangle$ are regarded as trail wave functions suggested in Refs.~\cite{liangSome1988,sandvikLoop2010} with higher projector performance to access real ground properties, $H=\sum_{i,j}\textbf{S}_{i}\cdot\textbf{S}_{j}$ is the antitferromagnetic Heisenberg model on a bipartite lattice (we set the interaction strength $J=1$ as the energy unit in this paper), and $m$ is the projection length. 
Following the standard PQMC procedure, one needs to rewrite the Hamiltonian as $H=-\sum_{p} H(p)$ with operator $H(p) = \frac{1}{4}-\textbf{S}_{1}(p)\cdot\textbf{S}_{2}(p)$, where $\textbf{S}$ represents spin, $1$ and $2$ are the two sites of bond $p$, belonging to two different sublattices of a bipartite system. The total configuration space of PQMC can be represent as a set $\{C\}=\{ V_l^0, p_1 \cdots p_m, V_r^0 \}$, and Eq.~\eqref{eq:swap} can be re-expressed as 
\begin{equation}\label{eq:projectorqmc2}
\langle \text{SWAP}_A \rangle = \frac{ \sum\limits_{\{C\}} w_{l} w_{r}\left\langle V_{l}^0\left|\prod\limits_{i=1}^m H(p_i) \text{SWAP}_A \prod\limits_{j=1}^m H(p_j) \right| V_r^0 \right\rangle}{  \sum\limits_{\{C\}} w_{l} w_{r} \left\langle  V_{l}^0 \left|\prod\limits_{i=1}^{2m} H(p_i) \right|V_{r}^0 \right\rangle}.
\end{equation}
We could sample the evaluation of SWAP operator in a particular configuration $C$ as
\begin{equation}
\text{SWAP}_A = \frac{\left\langle V_{l}^0\left|\prod\limits_{i=1}^m H(p_i) \text{SWAP}_A \prod\limits_{j=1}^m H(p_j) \right| V_r^0 \right\rangle}{\left\langle V_{l}^0\left|\prod\limits_{i=1}^{2m} H(p_i) \right| V_r^0 \right\rangle}
\end{equation}
in terms of the updating weight 
$ W_{C}= w_{l} w_{r} \left\langle  V_{l}^0 \left| \prod\limits_{i=1}^{2m} H(p_i) \right|V_{r}^0  \right\rangle
$; in other words,
\begin{equation}
\langle \text{SWAP}_A  \rangle = \frac{ \sum\limits_{C} W_{C}\  \text{SWAP}_A }{  \sum\limits_{C} W_{C} }.
\end{equation}
A combined VB-spin basis had been introduced in PQMC for fast local and global updating~\cite{sandvikLoop2010} by making $W_{l r p_1 \cdots p_m}$ to be independent to indices $p_1 \cdots p_m$, and giving rise to a less computational effort $O(m)$.
Typically, $m=20N$, here $N=L^d$ is the total system size at $d$ spatial dimension, is enough for simulating ground state properties.
For more details on the PQMC methodology and the definition of the SWAP operator, the readers are referred to Refs.~\cite{sandvikLoop2010,kallinAnomalies2011}.

\begin{figure}[htp!]
	\centering
	\includegraphics[width=1.0\columnwidth]{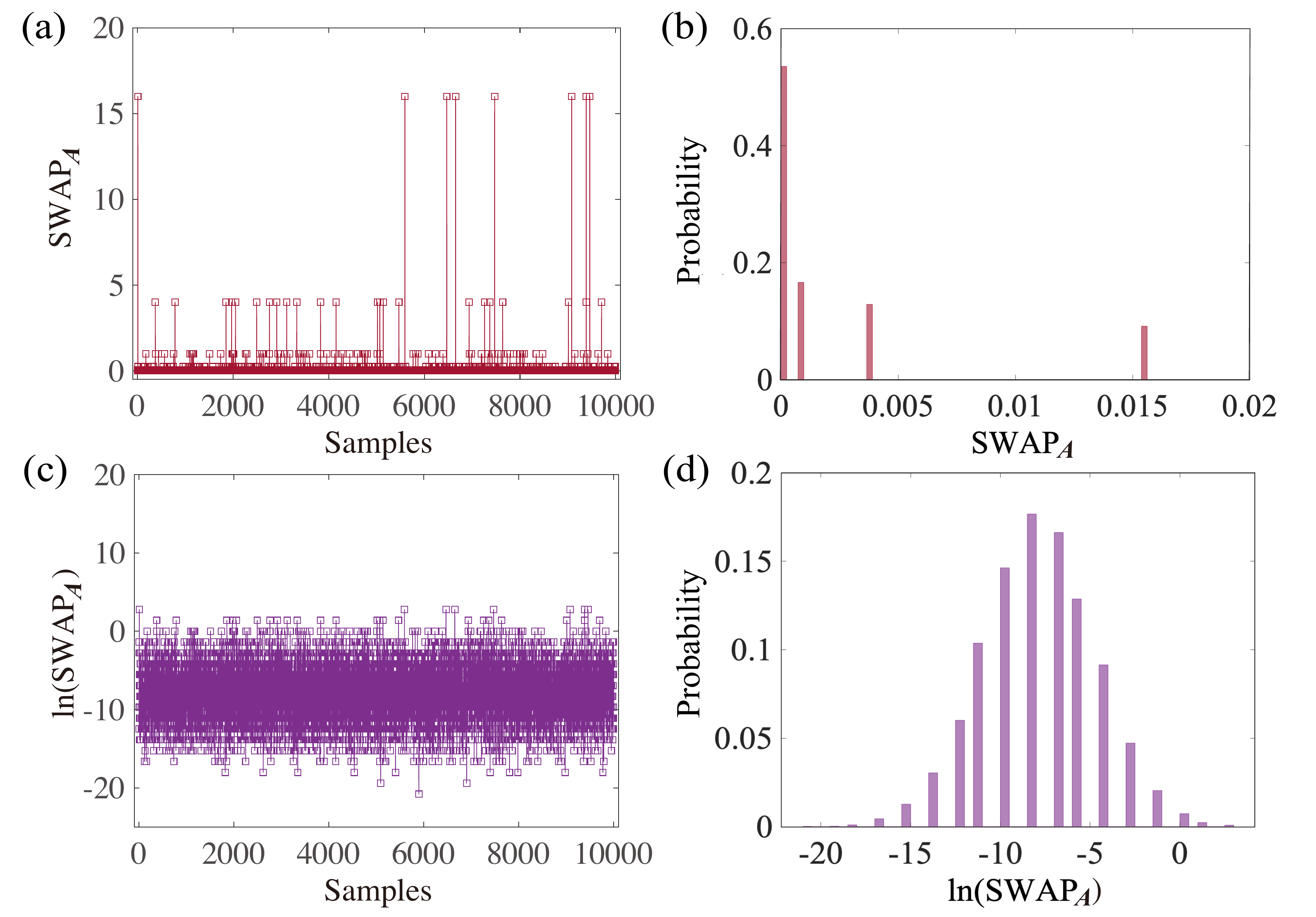}
	\caption{\textbf{SWAP operator is an exponential observable.} (a), (b) and (c), (d) show the sampling distributions and  normalized histograms of $\text{SWAP}_A$ and $\ln(\text{SWAP}_A)$ for a bipartition of Heisenberg model with smooth boundary (stripe region $A$ as in Fig.~\ref{fg:fig1} (a)) on a 10$\times$10 square lattice. $\text{SWAP}_A$ is obviously not normally distributed and the samples are mainly composed of very small values, with some sudden spikes from time to time. While $\ln(\text{SWAP}_A)$ is normally distributed , so it can be easily measured in QMC  simulation.}
	\label{fg:fig2}
\end{figure}

The $\langle \text{SWAP}_A \rangle=e^{-S_A^{(2)}}$ is an exponential observable~\cite{zhangIntegral2023}, and as $L$ increases, $S_A^{(2)} \propto L$ according to the area law, causing $e^{-S_A^{(2)}}$ to decrease exponentially, which makes estimating $\langle \text{SWAP}_A\rangle$ via PQMC extraordinarily challenging for large $L$. And this is the reason of the non-convergence problem in the previous SWAP operator calculation of EE. Below, we quantitatively elucidate such problem with numerical data. As depicted in Fig.~\ref{fg:fig2}, we plot the QMC sampling distribution and the corresponding histogram of a $10\times10$ lattice of 2d antiferromagnetic Heisenberg model with ``stripe'' subsystem for both $\text{SWAP}_A$ and $\ln{( \text{SWAP}_A )}$.
Evidently, as shown in Fig.~\ref{fg:fig2} (a) and (b), the vast majority of $\text{SWAP}_A$ samples lie extremely close to 0, markedly deviating from a normal distribution, which one can sample well in the QMC simulations. Strikingly, as shown in Fig.~\ref{fg:fig2} (c) and (d), the distribution of $\ln(\text{SWAP}_A)$ perfectly conforms to normality. 

\begin{figure}[htp!]
	\centering
	\includegraphics[width=1.0\columnwidth]{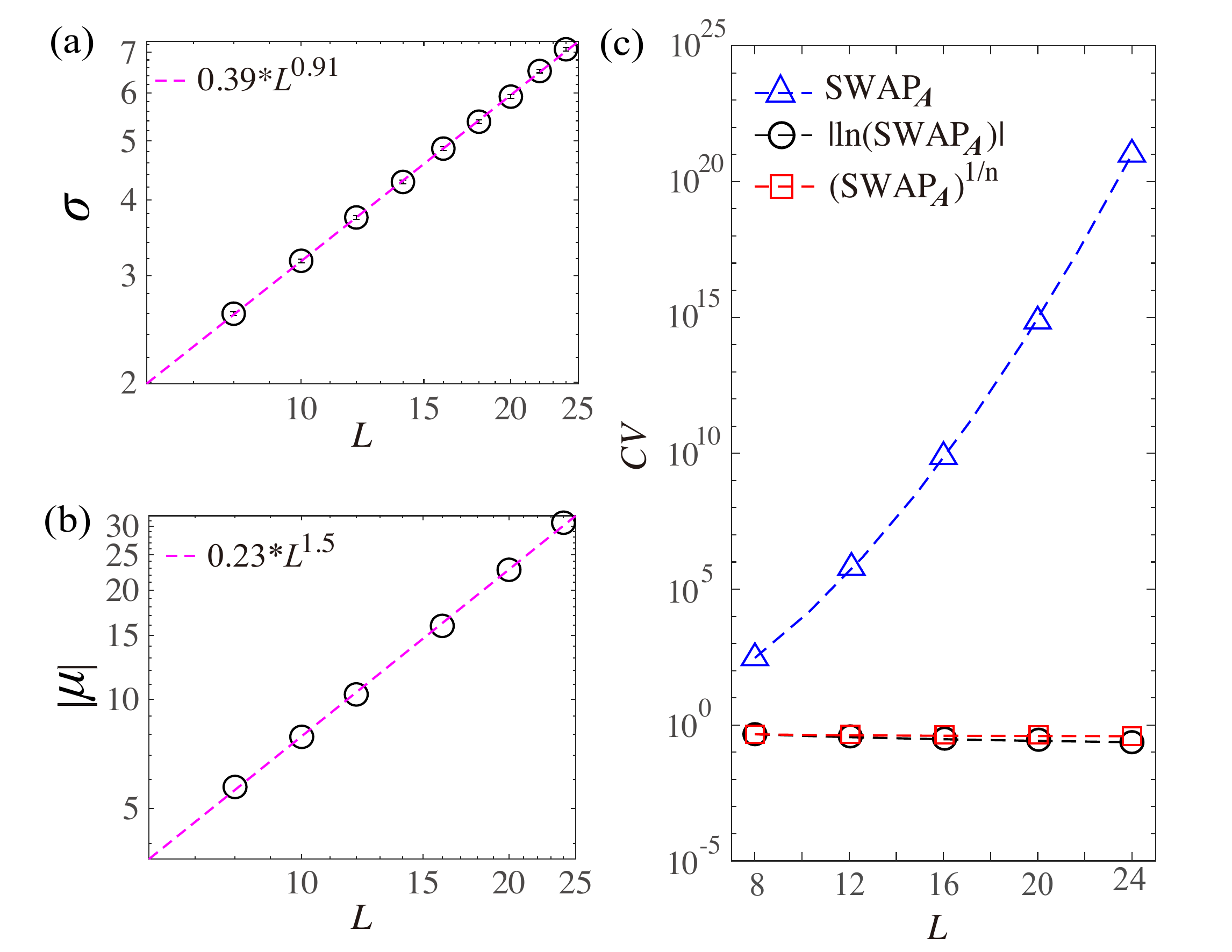}
	\caption{\textbf{Coefficient of variation of $\langle \text{SWAP}_A\rangle$ and its solution.} (a) and (b) show standard deviation $\sigma$ and expectation $\mu$ of $\ln(\text{SWAP}_A)$  for a bipartition of square lattice Heisenberg model with smooth boundary as a function of system size $L$ in the log-log scale. Both $\sigma$ and $\mu$ of $\ln(\text{SWAP}_A)$  increase  with $L$. The data can be fit with a  perfect power law as denoted by the dashed lines. (c) shows  $L$ dependence of coefficient of variation, CV, for three observables. Here, we choose $n$ as the ceiling integer of $|\langle\ln(\text{SWAP}_A)\rangle|$. CV[$\text{SWAP}_A$] increases exponentially while the others $\ln(\text{SWAP}_A)$ and $(\text{SWAP}_A)^{1/n}$ stay constant and even decreases with $L$, meaning that they are the ideal QMC observables with normal distribution.}
	\label{fg:fig3}
\end{figure}

We continue to quantify the explosion of the coefficient of variation of $\langle \text{SWAP}_A\rangle$.
Assume that $\ln(\text{SWAP}_A )$ obeys a normal distribution whose mean is $\mu$ and standard diviation is $\sigma$: $\ln(\text{SWAP}_A )\sim \mathcal{N}(\mu,\sigma^2)$.
The coefficient of variation of $\text{SWAP}_A$, given as the ratio of the standard deviation $\sigma_{\text{SWAP}_A}$ to the mean $\mu_{\text{SWAP}_A}$, can be computed as
\begin{equation}
    \text{CV}[\text{SWAP}_A] = \frac{\sigma_{\text{SWAP}_A}}{\vert\mu_{\text{SWAP}_A}\vert} = \sqrt{e^{\sigma^2}-1}.
\end{equation}
As shown in Fig.~\ref{fg:fig3} (a) and (b), we estimate the scaling relation of $\mu$ and $\sigma$ as function of subregion boundary length $L$ (as shown in Fig.~\ref{fg:fig1}, for both smooth boundary and corner cases, we design the boundary length of $A$ equal to twice of the linear size of the system $L$), which renders a perfect power-law. Thus, as show in Fig.~\ref{fg:fig3} (c), $\text{CV}[\text{SWAP}_A]$ exponentially explodes with $L$.

To address such non-convergence problem, we realize  $\text{CV}[(\text{SWAP}_A)^{1/n}]=\sqrt{e^{\sigma^2/n^2}-1}$, with $n$ to appropriately scale with $L$, which can be used to suppress the exponential growth of coefficient of variation. 
Here, we choose $n$ as the ceiling integer of $\vert\langle \ln( \text{SWAP}_A ) \rangle\vert$, then $\text{CV}[(\text{SWAP}_A)^{1/n}]$ would decay with $L$, enabling accurate evaluation. In practice, choosing an integer $n$ such that $\text{CV}[(\text{SWAP}_A)^{1/n}]$decreases or approaches a constant with $L$ is sufficient to guarantee the convergence. A smaller $\text{CV}$ means that we just need fewer samples to calculate entanglement entropy accurately. As shown in Fig.~\ref{fg:fig3} (c), $\text{CV}[\ln (\text{SWAP}_A)]$ and $\text{CV}[(\text{SWAP}_A)^{1/n}]$ indeed exhibit similar decaying behavior as $L$ increases. This proves $(\text{SWAP}_A)^{1/n}$ can be computed accurately in analogous to $\ln (\text{SWAP}_A)$, and the sampling of it will render normal distributions.

\section{Incremental SWAP operator Method}
\begin{figure}[htp!]
\centering
\includegraphics[width=1.0\columnwidth]{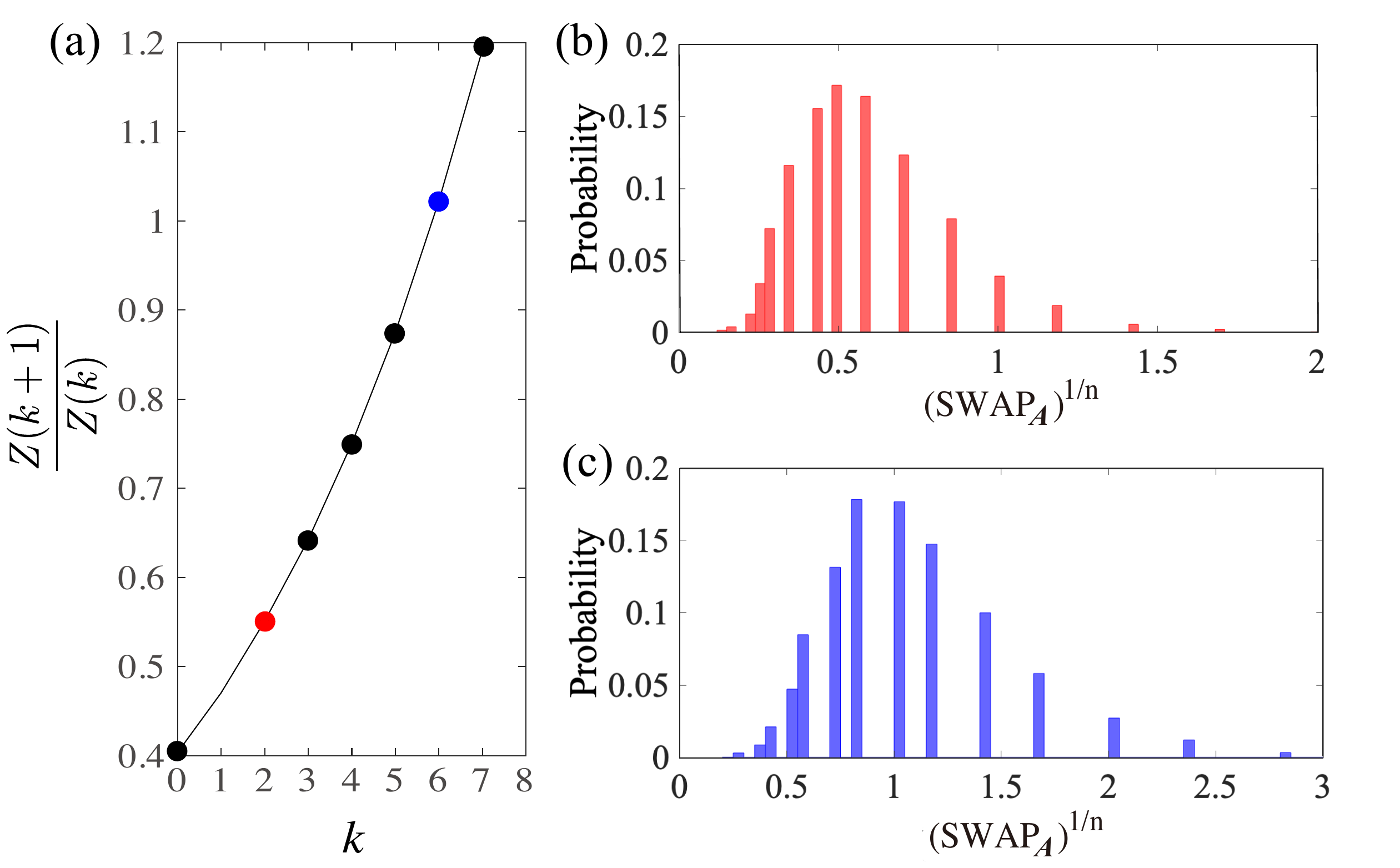}
\caption{\textbf{Incremental SWAP operator method.} (a) $\frac{Z(k+1)}{Z(k)}$ as a function of incremental process $k$ for a bipartition of $L=10$ square lattice Heisenberg model with smooth boundary and the total increments $n=8$ is  the ceiling integer of $\vert\ln (\text{SWAP}_A)\vert$. Factor $Z(k+1)/Z(k)$ is evaluated independently in each parallel process  to avoid introducing systematic errors. (b) and (c) show normalized histograms of $\text{(SWAP)}^{1/n}$ in $k=2$ (red dot in (a)) and $k=6$ (blue dot in (a)) incremental process. The normal distribution means that $\langle\text{(SWAP)}^{1/n}\rangle$ in each incremental process can be evaluated very accurately.}
	\label{fg:fig4}
\end{figure}

Based on the above analysis, we present an expression for our {\it incremental SWAP operator} method as follows
\begin{equation}\label{eq:incremental}
 	 \langle \text{SWAP}_A\rangle=\frac{Z(1)}{Z(0)}\frac{Z(2)}{Z(1)}\cdots\frac{Z({k+1})}{Z(k)} \cdots\frac{Z({n})}{Z({n} - 1)},
\end{equation}
where integer $k$ denotes the $k$-th increment with $n$ total increments, and we define the incremental partition function
\begin{equation}
Z\left(k\right)=\sum_{C} W_{C} ( \text{SWAP}_A) ^{k/n},
\end{equation}
with $Z(0) = \sum_{C} W_{C}$ and  $Z({n}) = \sum_{C} W_{C} \text{SWAP}_A$, naturally satisfying the 2nd R\'enyi EE definition.
Subsequently, each increment is evaluated in parallel per PQMC as follows
\begin{equation}\label{eq:incre-slice}
\frac{Z\left({k+1}\right)}{Z\left(k\right)}=\frac{\sum\limits_{C} W_{C}  ( \text{SWAP}_A )^{k/ n} ( \text{SWAP}_A)^{1/n} }{\sum\limits_{C} W_{C} ( \text{SWAP}_A )^{k/n}},
\end{equation}
where $n$ constitutes the ceiling integer of $\vert\mu\vert$. 
The new sampling observable is $( \text{SWAP}_A )^{1/ n}$ and the new sampling weight is $W_{C} ( \text{SWAP}_A )^{k/ n}$. 
Our incremental SWAP operator method allows us to evaluate the new observables $( \text{SWAP}_A )^{1/ n} $ for each piece of parallel incremental process with controlled statistical errors, as illustrated in Fig.~\ref{fg:fig4}(a). As expected, all pieces of the incremental process should generate reliable samples that approximately follow a normal distribution and have a finite variance. To demonstrate this, we have selected two parallel pieces and plotted their sample distributions and histograms in Fig.~\ref{fg:fig4} (b) and (c).

It is necessary to  discuss the effects of the accumulated error here. we simulated each factor in Eq. ~\eqref{eq:incre-slice} independently, which will give rise to a standard deviation factor $\sigma_i = e^{\mu/n^2 + \sigma^2 / 2n^2} \sqrt{e^{\sigma^2/n^2} - 1}$, and $\mu_i = e^{\mu/n^2 + \sigma^2 / 2n^2}$. According to the error propagation equation, the accumulated error for the product of $n$ factors in Eq. ~\eqref{eq:incre-slice} could be demonstrated with coefficient of variation $\text{CV}_{total}= \sqrt{ \sum_i^n  \sigma_i^2 /{\mu_i^2}} \sim \sigma/\sqrt{|\mu|}$. It is crucial to realize that our incremental SWAP operator method successfully reduces the total coefficient of variation from exponential growth to polynomial growth as $L$ increases.

Moreover, we also made improvement in the PQMC update scheme to reduce the computational complexity for SWAP operator calculation. We notcie the direct updating configurations according to Eq.~\eqref{eq:incre-slice}  now acquires an computational complex as $O(m^2+mN)$.
Typically, when sampling, one first transfers $\left\vert V_r^0\right\rangle$ to $\left\vert V_r^m\right\rangle=\prod\limits_{i=1}^{m} H(p_i)$ and $\left\langle V_l^0\right\vert$ to $\left\langle V_l^m\right\vert$, which requires $O(m)$ operations~\cite{sandvikGround2005,sandvikLoop2010,kallinAnomalies2011}. Then, sampling of $ \text{SWAP}_A = \left\langle V_{l}^m \left|\text{SWAP}_A \right| V_r^m \right\rangle / \left\langle V_l^m | V_r^m \right\rangle$ is performed, needing $O(N)$ operations to calculate the bra-ket overlap.
This huge computational burden prevents the efficient simulation from large system size.
To address this problem, we propose a new updating strategy called {\it propagating update}.

We notice that it is not necessary to restrain VB state at the center of projector time slice when sampling $\text{SWAP}_A$, in fact, one could perform such samplings at any projector time slice.
This property is guaranteed by the following equation, no matter what $j$-th time slice is,
\begin{equation}\label{eq:pqmc3}
\left\langle V_{l}^0\left|\prod\limits_{i=1}^{2m} H(p_i) \right| V_r^0 \right\rangle = 2^{-\left( o_l^j+o_r^{2m-j} \right)} \left\langle V_{l}^j \vert V_r^{2m-j} \right\rangle.
\end{equation}
Here $o_l^j$ and $o_r^{2m-j}$ are the numbers of operations (see Ref.~\cite{sandvikGround2005} for details) projecting from $V_l^0$ and $V_r^0$ to $j$-th projector time slice, respectively.
The same property is also valid to $\left\langle V_{l}^0\left|\prod\limits_{i=1}^m H(p_i) \text{SWAP}_A \prod\limits_{i=1}^m H(p_i) \right| V_r^0 \right\rangle$.
Thus the adjacent VB state can be updated locally, we don't need transfer $V_l^0$ and $V_r^0$ to $V_l^m$ and $V_r^m$ any more and the computational complex reduces from $O(m^2+mN)$ to $O(mN)$.
We further notice that the ratio of SWAP observable before and after update can be expressed as
\begin{equation}\label{pqmc5}
	\frac{(\text{SWAP}_A)^\prime}{\text{SWAP}_A} =\frac{ \frac{\left\langle V_{l}^0\left|\prod\limits_{i=1}^m H(p_i) \text{SWAP}_A \prod\limits_{i=1}^m H(p_i) \right| V_r^0 \right\rangle^\prime}{\left\langle V_{l}^0\left|\prod\limits_{i=1}^m H(p_i) \text{SWAP}_A \prod\limits_{i=1}^m H(p_i) \right| V_r^0 \right\rangle}}{  \frac{\left\langle V_{l}^0\left|\prod\limits_{i=1}^{2m} H(p_i) \right| V_r^0 \right\rangle^\prime}{\left\langle V_{l}^0\left|\prod\limits_{i=1}^{2m} H(p_i) \right| V_r^0 \right\rangle} }.
\end{equation}
According to Eq.~\eqref{eq:pqmc3}, the ratio of $\left\langle V_{l}^0\left|\prod\limits_{i=1}^{2m} H(p_i) \right| V_r^0 \right\rangle$ before and after update, for example at $j$-th projector time slice, is given by
\begin{equation}\label{eq:pqmc4}
\frac{\left\langle V_{l}^0\left|\prod\limits_{i=1}^{2m} H(p_i) \right| V_r^0 \right\rangle^\prime}{\left\langle V_{l}^0\left|\prod\limits_{i=1}^{2m} H(p_i) \right| V_r^0 \right\rangle} = 2^{-\left(o_l^{j\prime}- o_l^{j} \right)} \frac{\left\langle V_{l}^{j\prime} \vert V_r^{2m-j} \right\rangle}{\left\langle V_{l}^j \vert V_r^{2m-j} \right\rangle}.
\end{equation}
The effort to calculate this ratio in our propagating update is proportional to the average length of overlap loops of VB states~\cite{liangSome1988,sandvikLoop2010}, and the average length is proportional to the average squared sublattice magnetization~\cite{evertzloop2003a}.
For example, the effort to calculate Eq.~\eqref{eq:pqmc4} should be $O(N)$ in antiferromagnetic ordered phase and $O(1)$ in paramagnetic ordered phase or at the QCP.
The calculation of ratio for $\left\langle V_{l}^0\left|\prod\limits_{i=1}^m H(p_i) \text{SWAP}_A \prod\limits_{i=1}^m H(p_i) \right| V_r^0 \right\rangle$ before and after update has the same computational complexity.
It is notable that our novel propagating update strategy can be implemented using either the combined VB-spin basis~\cite{sandvikLoop2010} or the pure VB basis~\cite{sandvikGround2005}, while maintaining comparable computational complexity between the two.

In total, our {\it incremental SWAP operator} method with {\it propagating update} scheme for evaluating $S^{(2)}_A$ has a superior computational complexity of $O(mP)$ with $P = N$ or $1$ for magnetic or paramagnetic states. The full complexity follows as $O(mPL^{\gamma})$ given the requisite $n =\vert \ln (\text{SWAP}_A )\vert \sim L^{\gamma}$ increments. For Heisenberg model, $\vert \ln (\text{SWAP}_A) \vert = 0.23L^{1.5}$ in Fig.~\ref{fg:fig3} (b), thus $\gamma=1.5$.
Ascertaining $\gamma$ necessitates analysing $\mu(L)$ and $\sigma(L)$ scaling, costing $O(m)$. However, fitting merely requires a few small $L$. Therefore, this negligible overhead contrasts the total cost of evaluating EE. Up to now, the popular method to measure entanglement entropy with high efficiency and precision is non-equilibrium incremental algorithm \cite{demidioEntanglement2020,zhaoMeasuring2022}. 
  The computational complexity of this  method in PQMC is $m\times n_{qt}$, where $m$ is the projector length and $n_{qt}$ is the quenching time. And such incremental algorithm can overcome the convergence issue of entanglement entropy computation when the quenching time $n_{qt}$ is sufficiently long. However, a quantitative value for  $n_{qt}$ has not been determined yet, while the advantage of our method is that the convergence of the calculation can be guaranteed by  determining the total increments $n$ precisely.

\section{Results}
\begin{figure}[htp!]
\centering
\includegraphics[width=1.0\columnwidth]{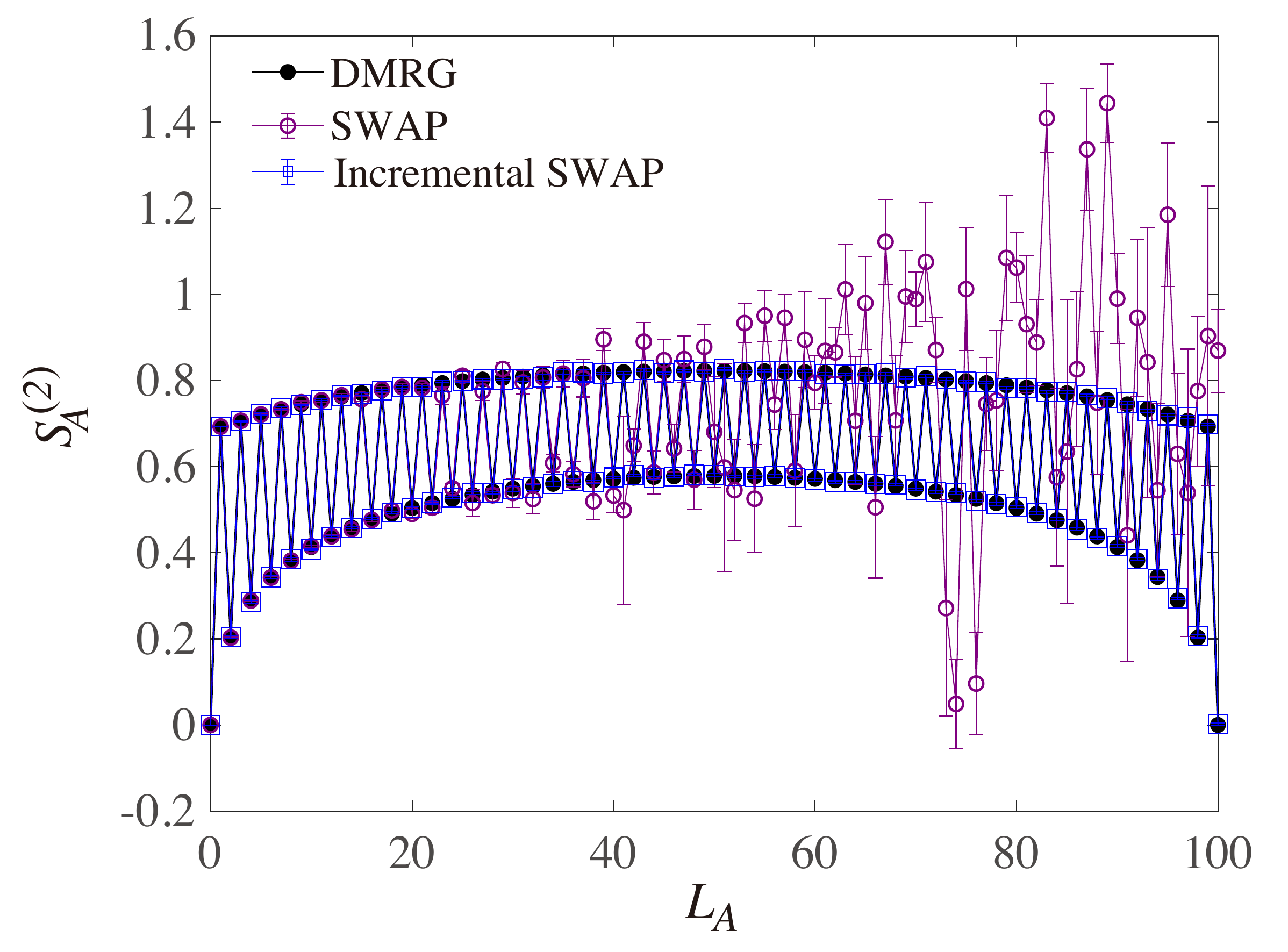}
\caption{\textbf{Incremental SWAP operator in 1d Heisenberg chain.} The 2nd R\'enyi entropy $S_A^{(2)}$ for the ground state of a $L=100$ Heisenberg chain with open boundary condition, as a function of subsystem size $L_A$. The black dots are DMRG results. Data labelled "SWAP" was calculated with the naive sampling as in Eq.~\eqref{eq:swap} 
%and in Refs.~\cite{hastingsMeasuring2010}
, while data labelled "Incremental SWAP" was calculated with our method in Eq.~\eqref{eq:incremental}, which is consistent with DMRG. The projection length $m/N = 50$ was used in both QMC simulations. "SWAP" and "Incremental SWAP" data consume the same CPU hours.}
	\label{fg:fig5}
\end{figure}

Even in 1d systems, directly employing the SWAP technique to compute EE encounters the non-convergence problem as the subregion increases~\cite{hastingsMeasuring2010}, our approach can overcome such problem. As shown in Fig.~\ref{fg:fig5}, our computed $S_A^{(2)}$ with incremental SWAP operator method for 1d Heisenberg chain with $L=100$ highly coincides with DMRG values. In contrast, $S_A^{(2)}$ computed directly via SWAP 
%as in Refs.~\cite{hastingsMeasuring2010,kallinAnomalies2011} 
grows increasingly unstable with expanding entanglement region size $L_A \ge L/2$. Besides the clear deviation of the mean values of  $S_A^{(2)}$, we also note that it is important to understand the errorbars of data obtained with the direct SWAP method are not credible, because the mean values of such data don't converge, due to the explosion of its coefficient of variation discussed in Fig.~\ref{fg:fig4}.

\begin{figure}[htp!]
\centering
\includegraphics[width=\columnwidth]{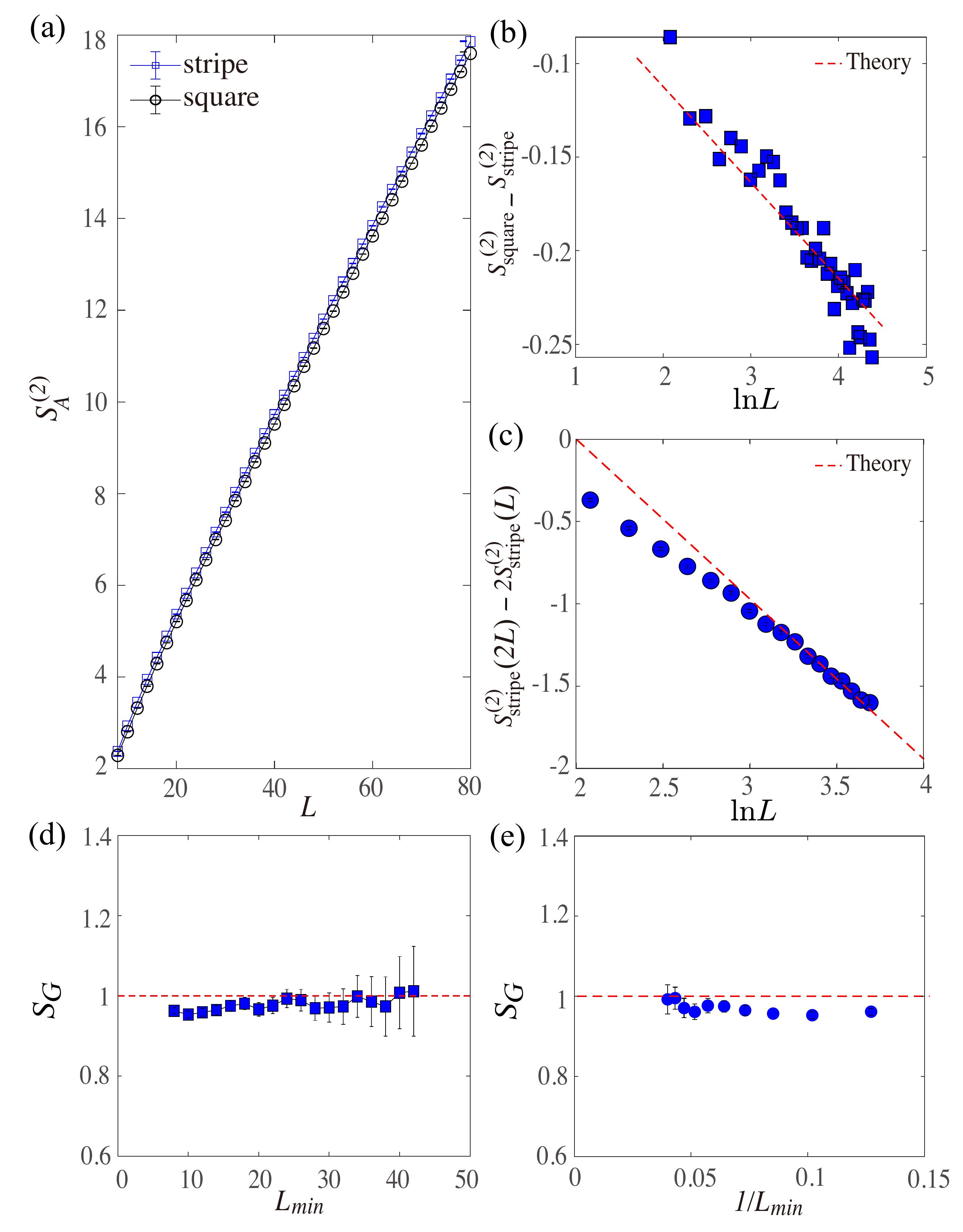}
\caption{\textbf{Incremental SWAP operator in 2d Heisenberg model.}  (a) The 2nd R\'enyi entropy $S_A^{(2)}$ for square lattice antiferromagnetic Heisenberg model as a function of system size $L$ for both the "stripe" (smooth boundary) and "square" (corner) entanglement regions $A$. The size of region $A$ is $L/2 \times L$ for stripe geometry and $L/2 \times L/2$ for square geometry.  Here we use projection length $m/N=20$ and the linear system size $L\in [8,80]$. (b) Comparison of QMC data to the theoretical prediction of $S_{\text{square}}^{(2)}-S_{\text{stripe}}^{(2)}$. Dashed line represents the theoretical prediction with a slope of $s_c=-0.0512$, symbols are data points. QMC data are coincidence with the spin wave calculation~\cite{laflorencieSpin-wave2015}. (c) Subtracted EE $S^{(2)}_{\text{stripe}}(2L)-2S_{\text{stripe}}^{(2)}(L)$ versus $\ln L$. The slope of the data line in (c) indicates the log-coefficient $s_G=1$ at large $L$. (d) The solid squares are the fitted $s_G$ according to Eq.~\eqref{eq:s2-stripe-deng} as $L_{min}$ increases. The dashed line represents $s_G=1$ for $n_G=2$ in the N\'eel state. $s_G$ robustly converges to 1. (e) The fitted $s_G$ from data in (c) for the subtracted EE with respect to the smallest retained system size $1/L_{min}$, the $s_G$ also robustly converges to 1.}
\label{fg:fig6}
\end{figure}

Next, we move on to the major numerical result of this paper -- the scaling of $S_A^{(2)}$ with $L$ for the ground state of 2d antiferromagnetic Heisenberg model, which is the well-known antiferromagnetic state with $n_G=2$ Goldstone modes~\cite{metlitskiEntanglement2011,laflorencieSpin-wave2015,songQuantum2023}.
As shown in Eq.~\eqref{eq:eq2}, such quantum modes will contribute a logarithmic correction to EE for a stripe entanglement region with smooth boundary, as depicted in Fig.~\ref{fg:fig1} (a), with $s_G=1$ in Eq.~\eqref{eq:eq2}. However, directly fitting  Eq.~\eqref{eq:eq2} to obtain $s_G$ via QMC is challenging, requiring numerical results with large system sizes and high precision~\cite{zhaoMeasuring2022,songExtracting2023}.
What's more, when the system sizes are not large enough, it turns out one needs to incorporate finite-size effects to reliably extract $s_G$ as following equation, 
\begin{equation}\label{eq:s2-stripe-deng}
S_{\text{stripe}}^{(2)}=a L+s_G \ln \left(\sqrt{I(L)\rho_s(L)} L\right)+\gamma_{\text {ord }},
\end{equation}
where $I(L)$ and $\rho_s(L)$ are the finite-size dependent magnetic susceptibility and spin stiffness, respectively~\cite{dengImproved2023}. In principle, $\sqrt{I(L)\rho_s(L)}$ should equal to $\frac{\rho_s}{v}$ when $L\rightarrow\infty$.

With the help of our incremental SWAP operator method, we obtain the accurate results of $S_\text{stripe}^{(2)}(L)$ up to $L=80$, as shown in Fig.~\ref{fg:fig6}(a).
We extract the corresponding area law coefficient $a$, the universal logarithmic corrections $s_G$ and the geometry constant $\gamma_\text{ord}$ via fitting Eq.~\eqref{eq:s2-stripe-deng} using a fitting window $[L_\text{min},80]$ and the values of $I(L)$ and $\rho_s(L)$ reported in Ref.~\cite{dengImproved2023}.
We plot $s_G$ as function of $L_\text{min}$ in Fig.~\ref{fg:fig6} (d), and find good agreement with the theoretic value $s_G=1$.
We further compare our fitting values of $a$, $s_G$ and $\gamma_\text{ord}$ with previous studies for a variety of methods, as listed in Table.~\ref{table:tab1}.

What's more, since the incremental SWAP method greatly reduced the computational complexity, we can also readily compute the  $S^{(2)}_\text{square}(L)$ up to $L=80$ for entanglement region ``square'' with four $\pi/2$ corners, as depicted in Fig.~\ref{fg:fig1}(b).
And we further try to extract the corners correction coefficient $s_c$ with following formula
\begin{equation}
S_{\text{square}}^{(2)}-S_{\text{stripe}}^{(2)}=s_{c} \ln(L).
\label{eq:eq15}
\end{equation}
Notably, such formula doesn't suffer from the finite-size effect as extracting $s_G$ via function fitting.
Our results, as shown in Fig.~\ref{fg:fig6} (b), give rise to an estimate $s_c=-0.061(3)$ for $L_\text{min}=8$, and $s_c=-0.059(3)$ for $L_\text{min}=10$, respectively.
The theoretical prediction value of $s_c$ approximate $-0.052$~\cite{metlitskiEntanglement2011,helmesUniversal2016}. The fitting results for different $L_\text{min}$ values for all the coefficients are summarized in Table \ref{table:tab2}. We note that, before the present work, there exists no previous numerical work that can give all the coefficients, including the area-law coefficient $a$, the universal log-coefficients $s_G$ and $s_c$, as well as the geometry constant $\gamma_{\text{ord}}$ in one set of QMC simulations.

\begin{table}[]
	\begin{tabular}{c|cccc}
		\hline \hline & \multicolumn{1}{c}{$a$} & \multicolumn{1}{c}{$s_G$} & \multicolumn{1}{c}{$\gamma_{\text{ord}}$} & \multicolumn{1}{c}{$s_{c}$} \\
		\hline Wave-function~\cite{metlitskiEntanglement2011} & $\times$ & 1 & 0.737 & -0.0496~\cite{casiniUniversal2007} \\
		\hline Series expansions~\cite{kallinAnomalies2011} & 0.188(2) & 1 & $\times$ & -0.0496 \\
		\hline Spin-wave~\cite{songEntanglement2011} & 0.191 & 0.92 & $\times$ & $\times$ \\
		\hline Spin-wave~\cite{laflorencieSpin-wave2015} & $0.190216(1)$ & 1 & $0.737$ & $-0.0512(8)$ \\
		\hline QMC~\cite{kallinAnomalies2011} & 0.19(1) & 0.74(2) & $\times$ & -0.1 \\
		\hline QMC~\cite{zhaoMeasuring2022} & 0.184(2) & 1.00(9) & $\times$ & $\times$ \\
		\hline QMC~\cite{dengImproved2023} & 0.1861(3) & 0.99(1) & 0.78(3) & $\times$ \\
		\hline This Work & $0.1865(4)$ & 0.98(2) & $0.770(8)$ & $ -0.067(4)$ \\
		\hline \hline
	\end{tabular}
	\caption{Summary of our fitting results and comparison with previous literature. Our results are taken at $L_{min}=18$ as shown in Tab.~\ref{table:tab2}.}
	\label{table:tab1}
\end{table}

\begin{table}[]
	\begin{tabular}{llllll}
		\hline \hline &\multicolumn{1}{c}{$L_{min}$} & \multicolumn{1}{c}{$a$} & \multicolumn{1}{c}{$s_G$} & \multicolumn{1}{c}{$\gamma_{\text{ord}}$} & \multicolumn{1}{c}{$s_c$} \\
        \hline & 8 &0.1869(2) & 0.962(7) &0.781(2)&-0.061(3)\\
               & 10 &0.1871(2)&0.953(8)&0.786(3)&-0.059(3)\\
               & 12 &0.1870(3)&0.958(9)&0.783(4)&-0.061(3)\\
               & 14 &0.1868(3)&0.96(1)&0.780(5)&-0.062(4)\\
               & 16 &0.1866(3)&0.98(1)&0.773(6)&-0.066(4)\\
               & 18 &0.1865(4) &0.98(2) &0.770(8) &-0.067(4)\\
        
		\hline \hline
	\end{tabular}
	\caption{Coefficients of the area law,the logarithmic terms, both $s_G$ (fitting from Eq.~\eqref{eq:s2-stripe-deng}) and $s_c$ (fitting from Eq.~\eqref{eq:eq15}) , and the geometric constant $\gamma_{\text{ord}}$.}
	\label{table:tab2}
\end{table}

Lastly, we follow the example of Ref.~\cite{songExtracting2023} to directly fit the $s_G$ in the smooth boundary case with the "subtracted EE", as $S^{s}(L)=S^{(2)}_{\text{stripe}}(2L) -2S^{(2)}_\text{stripe}(L)$, the $S^{s}(L)$ will remove the area-law contribution and promote the subleading log-term as the leading term. The subtracted EE versus $\ln(L)$ data are shown in Fig.~\ref{fg:fig6} (c) and indeed, one sees at large $L$, the $S^{s}(L)$ versus $\ln(L)$ is approaching a straight line with slope $-s_G=-1$. We also fit the $S^{s}(L)$ by gradually remove the small system sizes $L_{min}$, and one can see in Fig.~\ref{fg:fig6} (e), that the obtained $s_G$ is approaching 1 steadily.

\section{Discussion}
In this work, we have quantitatively explained the true reason of the non-convergence problem for the direct calculation of EE with SWAP operator method, and further developed an {\it incremental SWAP operator} method with {\it propagating} update in PQMC to offer a solution of the problem. Our methods enjoy both the conceptual clarity of the properly sampling the exponential observable, therefore overcome the explosion of the coefficient of the variation of EE in the original approach, as well as the reduced computational complexity in our new updated scheme in the PQMC simulations.

We use the 1d and 2d antiferromagnetic Heisenberg models to demonstrate the superior performance of our method, in particular, in the 2d case, for the first time, we successfully extract the area law coefficient, the universal logarithmic corrections, both $s_G$ and $s_c$, as well as the universal geometry constant, $\gamma_{\text{ord}}$, from our thence obtained EE data, and found them consistent with the theoretical predictions.

Looking forward, it is extremely easy to extend our computation to other interesting 2d highly entangled spin/boson quantum many-body systems, such as the scaling behavior of the EE in deconfined quantum critical point (DQCP) models~\cite{songDeconfined2023,zhaoScaling2022,songExtracting2023}, the SU(N) systems~\cite{songResummation-based2023} and topologically ordered quantum spin liquid models~\cite{zhaoMeasuring2022}, to reliably extract all the scaling coefficients therein. Moreover, the incremental SWAP method will also find its usage in the fermionic quantum many-body lattice models with the projector auxiliary field fermion QMC algorithms, and the initial step have already been taken by some of us~\cite{daliaoControllable2023}. With more controlled data quality and reduced computation burden, one can address the fundamental questions such as the scaling behavior of EE, as well as other exponential observables such as free energy~\cite{zhangIntegral2023}, in the interacting Dirac fermions and fermion DQCP models~\cite{liuFermion2023,daliaoteaching2023,liaoDiracI2022,liaoDiracII2022,liaoDiracIII2022,liaoDiracIII2022}, the symmetric mass generation models~\cite{liuDisorder2023}, and in the uncharted territory of interacting fermion surface, non-Fermi-liquid QCPs and topological orders and fractional quantum and anomalous Hall systems~\cite{shaoEntanglement2015,jiangMany2023,mishmashEntanglement2016,emeryQuantum2000,luFractional2024,kumarNeutral2022}.

\begin{acknowledgements}
X.Z. and Y.Q. acknowledge the support from National Key R\&D Program of China (Grant No.2022YFA1403400) and from NSFC (Grant No. 12374144).
Z.Y.M. and Y.D.L. acknowledge the support from the Research Grants Council (RGC) of Hong Kong Special Administrative Region of China (Project Nos. 17301721, AoE/P701/20, 17309822, C7037-22GF, 17302223), the
ANR/RGC Joint Research Scheme sponsored by RGC
of Hong Kong and French National Research Agency
(Project No. A\_HKU703/22), the GD-NSF (no.2022A1515011007) and
the HKU Seed Funding for Strategic Interdisciplinary Research. We thank the Beijng PARATERA Tech CO.,Ltd. (URL:
https://cloud.paratera.com) for providing HPC resources that
have contributed to the research results reported within this
paper.
\end{acknowledgements}

\bibliography{ref.bib}

\end{document}